\documentclass[reprint,aps,prb,superscriptaddress,amsmath,amssymb,floatfix]{revtex4-2}
\usepackage{graphicx}
\usepackage{gensymb}
\usepackage{booktabs}
\usepackage[hidelinks]{hyperref}

\newlength{\rolename}

\begin{document}

\title{Magnetic x-ray spectroscopy at the Fe $L_{2,3}$ edges of amorphous YIG}

\author{D. M. Cheshire}
\affiliation{School of Physics, Engineering and Technology, University of York, York YO10 5DD, United Kingdom}
\author{N. Gajurel}
\affiliation{School of Physics, Engineering and Technology, University of York, York YO10 5DD, United Kingdom}
\author{P. Bencok}
\affiliation{Diamond Light Source, Harwell Science and Innovation Campus, Chilton, Didcot OX11 0DE, United Kingdom}
\author{P. Steadman}
\affiliation{Diamond Light Source, Harwell Science and Innovation Campus, Chilton, Didcot OX11 0DE, United Kingdom}
\author{D. Gianolio}
\affiliation{Diamond Light Source, Harwell Science and Innovation Campus, Chilton, Didcot OX11 0DE, United Kingdom}
\author{G. Cibin}
\affiliation{Diamond Light Source, Harwell Science and Innovation Campus, Chilton, Didcot OX11 0DE, United Kingdom}
\author{S. A. Cavill}
\email{stuart.cavill@york.ac.uk}
\affiliation{School of Physics, Engineering and Technology, University of York, York YO10 5DD, United Kingdom}

\date{\today}

\begin{abstract}
Spectroscopic magnetometry offered by x-ray magnetic circular dichroism (XMCD) has been performed on amorphous yttrium iron garnet (a-YIG) thin films to investigate the magnetic contribution of the two Fe$^{3+}$ sites, tetrahedral (T$_{d}$) and octahedral (O$_{h}$). The magnetic environment of the amorphous material is shown to be different to that in the crystalline state. The inverted XMCD lineshape indicates that a-YIG magnetism is dominated by O$_{h}$-coordinated Fe species, in contrast to crystalline YIG, for which atomic multiplet calculations using CTM4XAS give a T$_{d}$:O$_{h}$ ratio approaching the expected 3:2. O$_{h}$ dominated Fe magnetism is shown to be weak and stable with thermal annealing, up to at least 400~\celsius{}, changing significantly only upon recrystallisation of the YIG (650~\celsius{}), where T$_{d}$ Fe$^{3+}$ dominates over O$_{h}$ Fe species, as evidenced by XMCD. On cooling to 10~K, an additional, strongly temperature-dependent dichroic signal emerges on the low-energy side of the Fe $L_3$ edge. We attribute this to a minority, Curie-paramagnetic population of magnetically decoupled Fe$^{2+}$ ions at strongly distorted, under-coordinated sites. Atomic multiplet calculations for Fe$^{2+}$ in tetragonal symmetry reproduce the lineshape of this signal. Bulk-sensitive Fe K-edge x-ray absorption near-edge structure (XANES) shows an Fe$^{3+}$/$\Sigma$Fe ratio close to 1, confining Fe$^{2+}$ to a small fraction of the total Fe. This minority nevertheless dominates the low-temperature XMCD because, unlike the largely compensated Fe$^{3+}$ network, it is strongly polarised at low temperature.
\end{abstract}

\keywords{amorphous yttrium iron garnet, x-ray magnetic circular dichroism, x-ray absorption spectroscopy, XANES, magnetic insulators, multiplet calculations}

\maketitle

\section{Introduction}
The need to improve energy efficiency in modern ICT has  led to a focus on spin insulatronics research: the area of spintronics devoted to the transport and processing of information via spin waves in magnetic insulators \cite{Chumak2022, Flebus2024}. Due to its exceptionally low damping, even in thin films, Yttrium Iron Garnet (YIG) is considered the most prominent material in this field, being widely used in spin pumping experiments and research on magnonics \cite{Cornelissen20151022, Onbasli2014}. New research in this area has been stimulated by reports of relatively long spin decay lengths in amorphous YIG (a-YIG) \cite{Cheshire2024, Yu2024} where there is no significant long range structural or ferrimagnetic order (such that magnetisation is approximately zero at B = 0 T). These length scales are comparable to thin film antiferromagnets such as crystalline NiO and Cr$_2$O$_3$ \cite{Wang2015}, indicating that a-YIG may be used as a good transporter of angular momentum where thin insulators / semiconductors are needed rather than low spin-orbit coupled normal metals.

Spin transport through these disordered magnetic insulators is thought to rely only on the magnetic correlation length scales with respect to the grain size \cite{Ochoa2018}.
However, non-local experiments have failed to observe long range ($>$ 1 micron) spin transport through a-YIG \cite{Yang2021, GomezPerez2020}, or have produced a range of spin decay lengths (3 - 16 nm) for a-YIG via FMR spin-pumping experiments \cite{Wang2015, Cheshire2024, Yu2024}. This variation in the spin decay length gives weight to the magnetic correlation hypothesis for the spin transport mechanism in a-YIG, which is highly dependent on short range magnetic correlations. However, the nature of spin transport in a-YIG currently remains an area of open debate. One factor contributing to the uncertainty surrounding this material is that detailed structural and spectroscopic information on a-YIG is lacking. This gap limits our understanding of the fundamental properties of a-YIG, and whether these properties vary from sample to sample. Literature suggests the Fe environment in the amorphous material is different to that in the crystalline state, although existing reports disagree on this issue \cite{Gyorgy1979, Chukalkin1989TheY3Fe5O12}.

It is notable that studies of spin transport in a-YIG use different deposition techniques to fabricate the material under investigation. These include sputtering \cite{Wesenberg2017,Yang2021,GomezPerez2020,Wang2015} and PLD \cite{Cheshire2024}. The diversity of a-YIG fabrication methods also extends to much of the existing literature of ‘disordered YIG’ published several decades ago, that were prepared using either roller quenching, anodic oxidation or neutron irradiation \cite{Gyorgy1979, Simpson1971, Chukalkin1983}. This raises the possibility that samples of a-YIG may, in fact, have differing ‘amorphousness’ and Fe cation distributions or environments that are highly dependent on growth conditions. These differences are likely to contribute to variations in the reported behaviours. As a-YIG has M $\approx$ 0 emu/cc and no long range structural order, the usual techniques of vibrating sample magnetometry (VSM) and x-ray diffraction (XRD) provide little to no information. Even nano-diffraction in cross-sectional TEM shows no structural Bragg peaks \cite{Cheshire2024}. VSM runs into a bigger problem in that the preferred substrate for YIG growth, gadolinium gallium garnet (GGG) with a volume of order $10^4$ times larger than the thin film, is a Curie paramagnet ($\chi \propto \frac{1}{T}$) and so dominates the measured VSM signal, especially at low temperature. Spectroscopic magnetometry, offered by XMCD, is therefore essential to the study of local magnetism at the two Fe$^{3+}$ sites in a-YIG in order to provide atomic level information. To this extent, recent work has examined spin transport and XMCD measured in RF-sputtered a-YIG on GGG, both as-deposited and ex-situ annealed, at temperatures approaching the recrystallisation temperature of the YIG (T $\geq$ 550~\celsius{}). However, the authors conclude that a-YIG samples annealed below 580~\celsius{} yield almost zero XMCD signal when measured at 0.5 T, consistent with no magnetisation \cite{Yu2024}.

In this paper, we explicitly examine the Fe environment in the amorphous YIG phase using x-ray spectroscopy. A series of a-YIG thin films on YAG are characterised using normal incidence Fe $L_{2,3}$ x-ray absorption spectroscopy (XAS) and XMCD to probe the magnetism of the a-YIG Fe sites, and ascertain whether low temperature ex-situ annealing modifies this. Fe K-edge x-ray absorption near-edge structure (XANES) has also been performed to provide complementary measurements of the Fe oxidation state.

\section{Methods}
Amorphous YIG films were grown via pulsed laser deposition (PLD) on (111)-oriented yttrium aluminium garnet (YAG) substrates. YAG was chosen as the substrate because it is a diamagnetic garnet with a far smaller magnetic background than GGG, which makes VSM measurements less problematic. Furthermore, the recrystallisation dynamics of thin-film a-YIG deposited on YAG substrates are comparable to that on GGG substrates: with both initiating around 600~\celsius{} \cite{Yu2024,SailerPhysRevMaterials.8.043402}. This is unlike other diamagnetic substrates such as silicon or SiO\textsubscript{x}, with higher thermal budgets above 700-800~\celsius{} being required \cite{Gage2017Si,Zhang7120983,WangCO2:24}.

Following \cite{Hauser2016, Cheshire2024}, an amorphous a-YIG layer was deposited at room temperature, in a partial oxygen pressure of 2.5$\times 10^{-3}$ mbar, by ablating a stoichiometric polycrystalline YIG target. Ablation was performed using a frequency quadrupled Nd:YAG laser ($\lambda$ = 266 nm) with a fluence of approximately 1 J\,cm$^{-2}$, at a repetition frequency of 10 Hz. The sample was rotated at 6 RPM during deposition. This technique was used to produce several a-YIG films with thicknesses of (80 $\pm$ 3) nm, as determined by XRR.
Following deposition, the a-YIG-on-YAG films were ex-situ annealed in atmospheric conditions for 3 hours, with each sample at a different respective temperature. The first sample was annealed at 650~\celsius{} to achieve long range crystalline order, as a YIG-on-YAG control. The other a-YIG/YAG samples were left as-grown, or annealed at 200~\celsius{}, 400~\celsius{}, 500~\celsius{}, 550~\celsius{} and 600~\celsius{}. Films annealed at or below 550~\celsius{} remained amorphous, whereas the 600~\celsius{} film had begun to recrystallise (see Results). Not every film was used for every measurement: VSM data (figure~\ref{fig:vsm}(a)) were taken on the as-grown, 200, 400, 500 and 650~\celsius{} films, and Fe K-edge XANES (figure~\ref{fig:xanes}) on the as-grown, 400, 500, 550, 600 and 650~\celsius{} films. The temperature of each film was ramped at approximately 4 K\,min$^{-1}$. After annealing, a 2~nm platinum capping layer was deposited on the as-grown, 200~\celsius{}, 400~\celsius{} and crystalline (650~\celsius{}) samples to allow total electron yield (TEY) detection for XAS/XMCD spectroscopy measurements. The Pt layer was deposited at room-temperature using PLD, ablating a pure Pt target in vacuum (base pressure of 1$\times$10$^{-8}$ mbar).

After growth, VSM measurements were performed at room temperature on the a-YIG/YAG with the field applied in the plane of the sample.
Fe $L_{2,3}$ x-ray absorption spectroscopy (XAS) and XMCD measurements were performed on beamline I10 (BLADE) at the Diamond Light Source \cite{2016I10BLADE}. The XAS was measured in normal incidence by TEY using the sample drain current. In $L$-edge XAS / XMCD measurements, electrons are excited from a 2p core level to the unoccupied 3d valence states of the element of interest by circularly polarised x-rays at the resonance energy of the transition. The difference in absorption between opposite circular polarisations (XMCD) gives a direct and element-specific measurement of the projection of the 3d magnetic moment along the x-ray helicity vector \cite{vanderLaan2014X-rayMagnetism, XMCD}. The probing depth for TEY detection is approximately 6 nm at the $L_{2,3}$ absorption edges of 3d transition metals. All Fe $L_{2,3}$ XAS measurements were performed at normal incidence to minimise electron-yield saturation effects  on the spectra \cite{Nakajima1999} with a 5 T field applied collinearly to the x-ray helicity vector. Measurements were taken at temperatures of 10 K, 50 K, 100 K, 200 K, and 300 K. Additional hard x-ray absorption (XANES) was performed on the core XAS beamline B18 at the Diamond Light Source \cite{Dent2009B18:Diamond}. XANES was measured over the Fe K-edge to provide complementary information on the valence of the Fe cations. Both XMCD and XANES measurements were performed on the a-YIG/YAG samples, and compared with previously reported data for a crystalline YIG/GGG film \cite{Cheshire2022}.

\section{Results}

\begin{figure}[htbp]
    \centering
    \includegraphics[width=\linewidth]{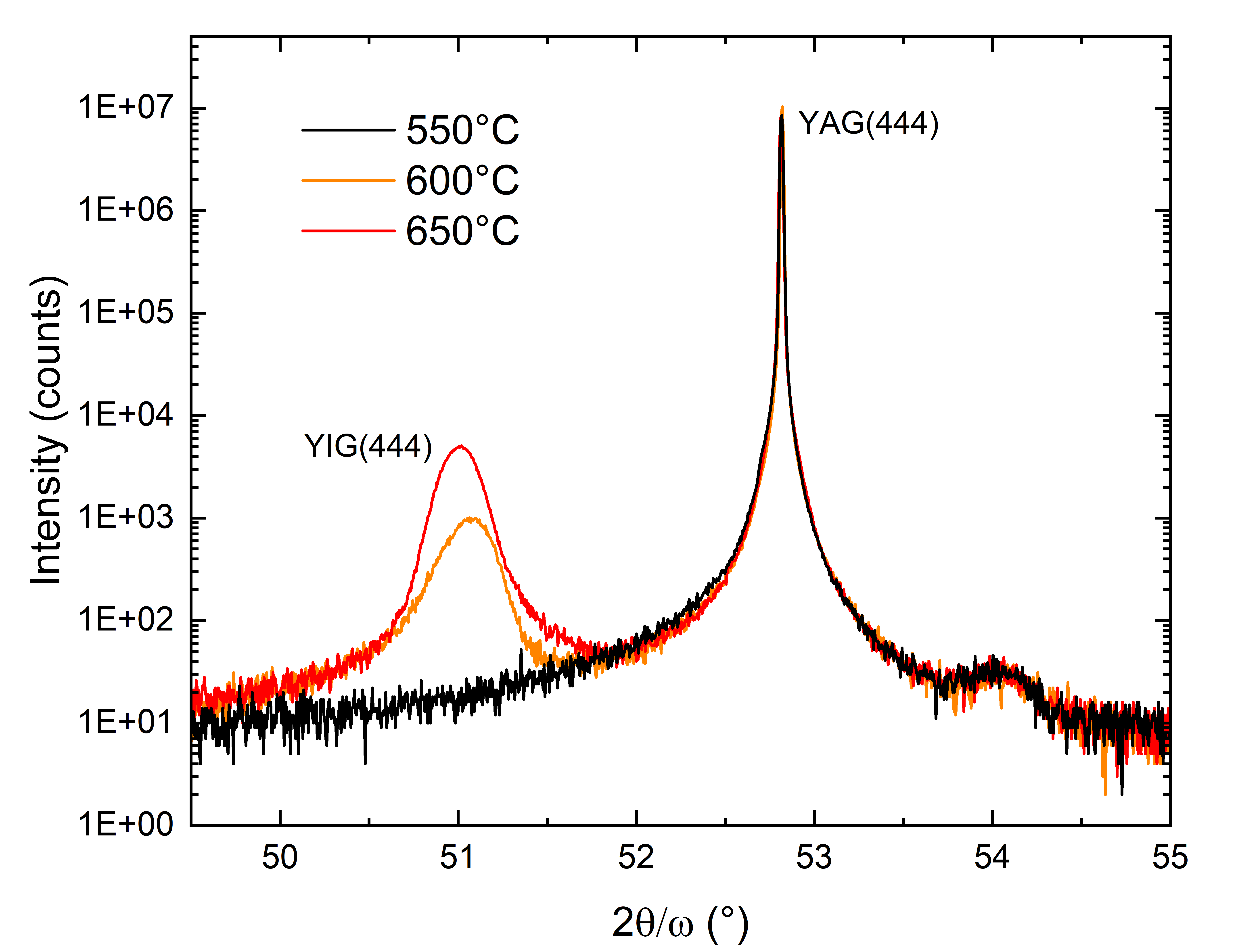}
    \caption{2$\theta$-$\omega$ XRD measurements for a-YIG/YAG(111) and YIG/YAG(111) films with different ex-situ annealing temperatures. Data for all films annealed at or below 550~\celsius{} are identical.}
    \label{fig:xrd}
\end{figure}

X-ray diffraction confirmed no obvious crystallisation of the a-YIG layers had occurred on the YAG substrate, following either deposition or ex-situ annealing at T $<$ 600~\celsius{}. 2$\theta$-$\omega$ scans, shown in figure~\ref{fig:xrd}, illustrate that for both as-deposited a-YIG and samples annealed below 600~\celsius{}, only the YAG(444) substrate reflection is observed. This confirms that a-YIG remains sufficiently amorphous in this temperature range to prevent coherent diffraction of x-rays. At 600~\celsius{} and 650~\celsius{}, an epitaxial YIG(444) reflection is observed in addition to the YAG(444) substrate. Epitaxial growth was confirmed with no other features, other than the family of allowed (nnn) reflections (where n is a multiple of 4), being present in broader 2$\theta$-$\omega$ scans. A weak, broad feature near 54\degree\ is present for all samples, including the amorphous films, and therefore originates from the substrate or instrument rather than the film. A clear shift in 2$\theta$-$\omega$ of $\approx$ 0.05\degree\ is observed between YIG(444) annealed at 600~\celsius{} and 650~\celsius{}.

The YIG plane spacing along the growth direction, d\textsubscript{444}, was calculated from Bragg's law (Cu K$\alpha_1$, with peak positions referenced to the YAG(444) substrate reflection at 2$\theta$ = 52.787\degree, $a_{\rm YAG}$ = 12.005~\AA) as (1.788 $\pm$ 0.001)\AA\ for 600~\celsius{} and (1.790 $\pm$ 0.001)\AA\ for 650~\celsius{} respectively, at most 0.2\% larger compared to bulk YIG (d\textsubscript{444} = 1.786 \AA).

\begin{figure}[htbp]
    \centering
    \includegraphics[width=\linewidth]{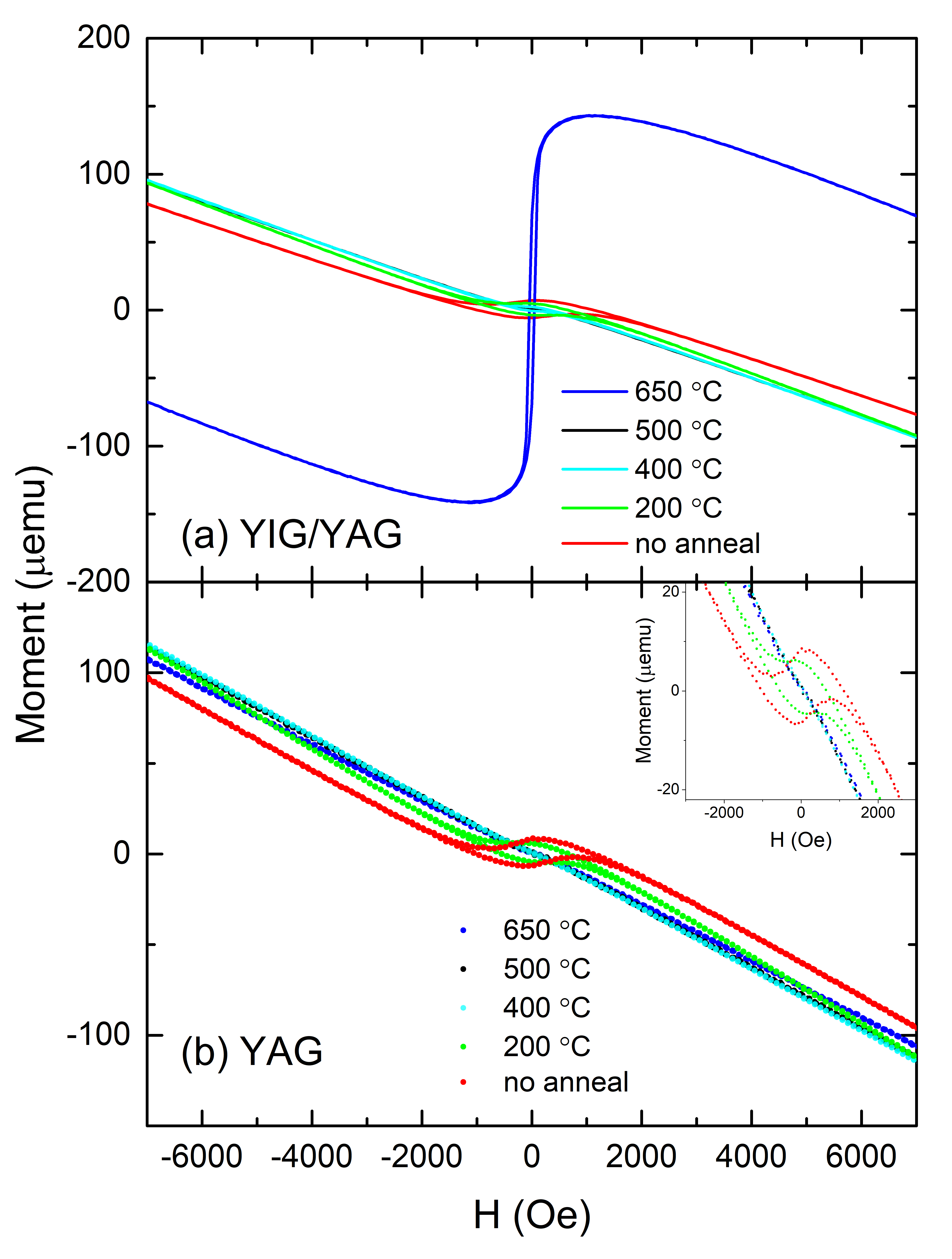}
    \caption{(a) Magnetic moment versus field (m-H) measurement of the YIG/YAG system at various annealing temperatures. (b) m-H measurement of the YAG substrates at various annealing temperatures. The inset shows details of the low applied field region.}
    \label{fig:vsm}
\end{figure}

VSM measurements were recorded for the as-deposited and annealed YIG/YAG samples. After annealing at 650~\celsius{}, long range magnetic order is established as evidenced by the increase in saturation and remanent magnetisation, as shown in figure~\ref{fig:vsm}(a), that correlates to the structural order. However, the results also demonstrate an unexpected opening of the m-H loop for the as-grown (no anneal) sample that gradually disappears upon annealing to temperatures around 400~\celsius{}, contrary to the expected paramagnet on diamagnetic substrate behaviour. The cause of the loop opening and change in magnetic susceptibility upon annealing becomes apparent when measuring m-H of the YAG substrates before growth of the a-YIG films.

\begin{figure}[htbp]
    \centering
    \includegraphics[width=\linewidth]{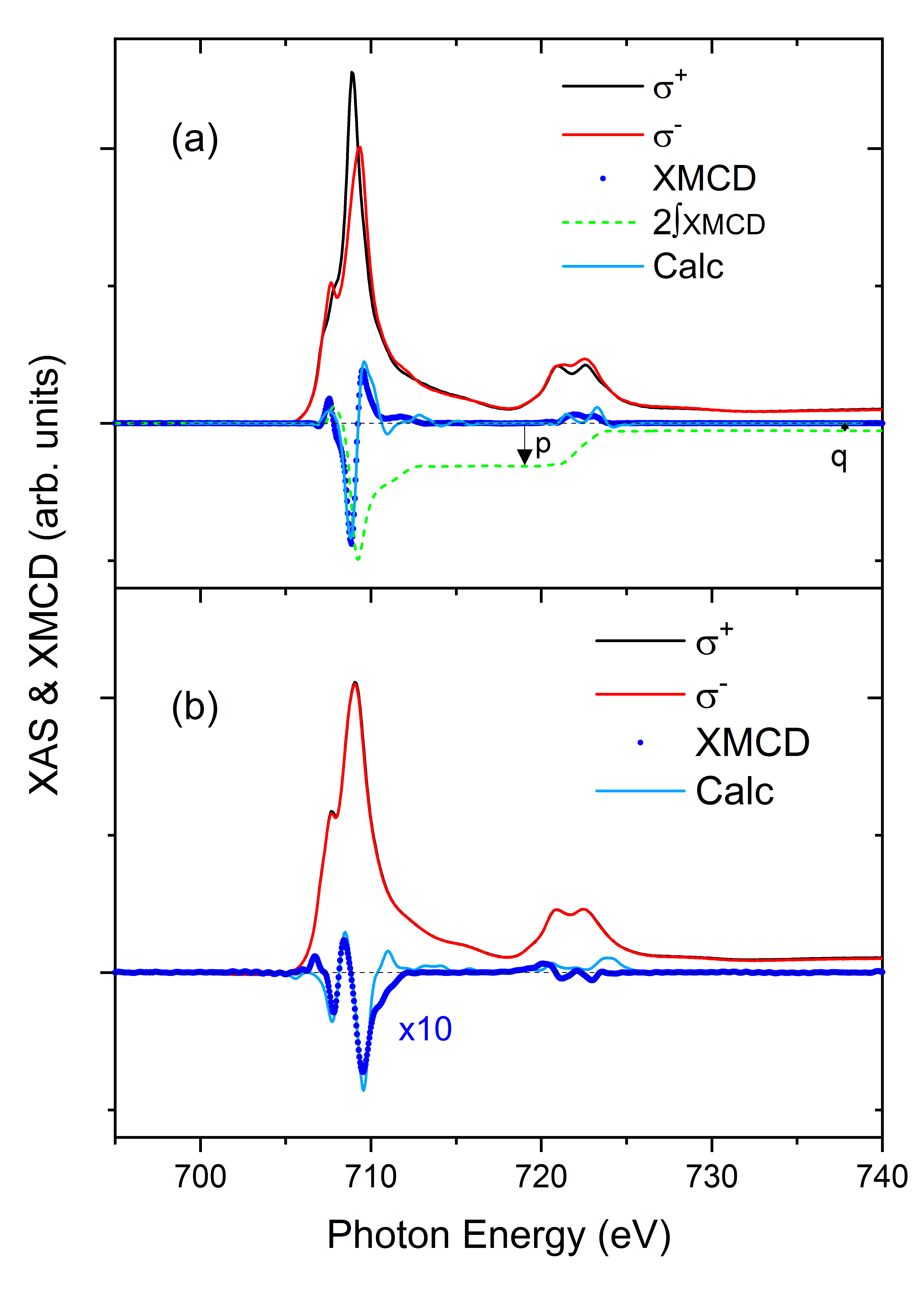}     
\caption{Experimental XAS/XMCD spectra for (a) 80~nm YIG/YAG thin film, and (b) 80~nm a-YIG/YAG thin film. The spectra were measured using TEY at normal incidence in a 5~T magnetic field at 300 K. The XMCD spectra from atomic multiplet calculations are also shown.}
\label{fig:xmcd300}
\end{figure}

Figure~\ref{fig:vsm}(b) shows m-H data for the YAG substrates (directly out of manufacturer packaging) and annealed in air at various temperatures. Data was also taken for samples ultrasonically cleaned in acetone and IPA which showed the same trend as in figure~\ref{fig:vsm}(b) (not shown). For the unannealed sample, a clear opening in the m-H loop is observed along with slight non-linearities in the magnetic susceptibility at fields above 3000 Oe. As the substrates are annealed in air to progressively higher temperatures, the loop opening closes and the linearity improves, similar to the YIG/YAG system for temperatures below 500~\celsius{}. A susceptibility for YAG of $-$1.2 $\times$10$^{-6}$ emu cm$^{-3}$ Oe$^{-1}$ (CGS volume susceptibility) was obtained from the $T_{anneal}$$ > $ 400~\celsius{} data.  Differences in the susceptibility between $T_{anneal}$ = 500~\celsius{} and 650~\celsius{} are of the order of 6\% which is similar to the variation in substrate volume. The magnetic signature of the YAG substrates is likely due to oxygen defects that are removed when annealed above 400~\celsius{} in air. However, this unexpected magnetic response of the substrate and the variation in the moment from the volume uncertainty makes removal of the background signal from the a-YIG/YAG data highly error prone. Therefore we use the element specific nature of XMCD to probe the Fe magnetism of the film directly without substrate contributions.
XMCD spectra for the YIG/YAG thin film are shown in figure~\ref{fig:xmcd300}(a). The XAS / XMCD lineshape is similar to that of YIG thin films grown on GGG substrates \cite{Cheshire2022, Vasili2017DirectFilms, Krichevtsov2017}. Three peaks exist in  the XMCD, spanning the Fe L$_{3}$ edge (707-710 eV): a negative peak at 708.8 eV (the most intense), a strong positive peak at 709.5 eV and a weaker positive peak at 707.6 eV. The two most intense peaks are due to magnetic spectral features associated with the T$_{d}$ sites and O$_h$ sites respectively, that are aligned anti-parallel to each other in the ferrimagnetic structure. The less intense positive peak at the lower energy is part of the O$_h$ multiplet structure.

Integration of the YIG XMCD over the L edges gives $p = -0.040 \pm 0.002$ and $q = -0.006 \pm 0.002$, where p and q are the integrated intensities over the L$_3$ and L$_{2,3}$ XMCD signal, respectively, seen in figure~\ref{fig:xmcd300}(a). Using the XMCD sum rules, $\mu$\textsubscript{L}/$\mu$\textsubscript{S} $= 2q/(9p-6q)$ $= 0.037 \pm 0.015$ was calculated, corresponding to a g-factor, $g = 2(1+\mu_L/\mu_S)$, of $2.07 \pm 0.03$. The g-factor is close to 2, but slightly outside the value reported for YIG/GGG thin films \cite{Cheshire2022}. Because high-spin Fe$^{3+}$ ($3d^5$) nominally has $L = 0$, the finite orbital contribution most likely reflects systematic uncertainties (i.e $j-j$ mixing, finite background) in the sum-rule analysis rather than a true large orbital moment 

In contrast, the XMCD spectrum for the as-deposited a-YIG/YAG thin film, shown in figure~\ref{fig:xmcd300}(b), is very different. The magnitude of the a-YIG XMCD is an order of magnitude smaller than the crystalline YIG and vanishes in zero applied field \cite{Yu2024}. The most significant dichroism is observed over the L$_{3}$ edge; dichroism over L$_{2}$ is minimal. The a-YIG XMCD lineshape at the L$_{3}$ edge shows three distinct peaks, but with orientations opposite to those of YIG; a positive peak at 708.4 eV, and two negative peaks at 707.8 eV and 709.5 eV (the latter being the strongest). At these moderate fields (B = 5 T) the moments are unlikely to be fully saturated \cite{GomezPerez2020} and therefore we do not extract absolute moments from sum rule analysis. The ratio of the integrated $L_2$ and $L_3$ dichroism, which for a given species does not depend on the degree of alignment, is used below only to compare orbital contributions.

Contributions to YIG/YAG magnetism from tetrahedral and octahedral Fe$^{3+}$ sites were evaluated by atomic multiplet calculations using the CTM4XAS software \cite{Stavitski2010}. A Slater parameter reduction of 80\%, 80\%, 80\% was used, with 10Dq values of 1.6 eV and $-$0.6 eV for O$_{h}$ and T$_{d}$ sites respectively, following \cite{Kuiper1997FeFe3O4}. A Lorentzian width of 0.15 eV was used for the L$_{3}$ edge and 0.3 eV for the L$_{2}$, with Gaussian broadening of approximately 0.2 eV across the whole energy range. The atomic multiplet calculation, shown in figure~\ref{fig:xmcd300}(a), agrees well with the experimental XMCD data from the YIG/YAG film at the L$_{3}$ edge for a relative occupation of Fe sites (Fe$^{3+}$ T$_{d}$: Fe$^{3+}$ O$_{h}$) of 62:38, with a 5\% tolerance on each. This is within error the expected ratio of T$_{d}$:O$_{h}$ Fe$^{3+}$ in stoichiometric YIG of 3:2, with no Fe$^{2+}$ contributions.

On the other hand, atomic multiplet calculations suggest that the magnetism in a-YIG is dominated by O$_{h}$ coordinated Fe species, in contrast to the crystalline YIG which is dominated by T$_{d}$ 
coordinated Fe$^{3+}$. This dominance of O$_{h}$ coordinated Fe is shown in the reversal of the XMCD signal in a-YIG, where the O$_{h}$ and  T$_{d}$ XMCD peaks have reversed sign compared to figure~\ref{fig:xmcd300}(a). Multiplet calculations using the same input parameters as for the crystalline YIG fail to reproduce the XMCD lineshape, even when allowing the O$_h$ to T$_d$ ratio to vary. A slightly better fit is obtained with an Fe$^{2+}$ O$_{h}$ contribution, as shown in figure~\ref{fig:xmcd300}(b). If present, Fe\textsuperscript{2+} at the a-YIG surface would allow hopping conduction between Fe\textsuperscript{2+} and Fe\textsuperscript{3+} O$_h$ sites, which has been proposed to explain non-local signals in a-YIG \cite{GomezPerez2020} and its low room-temperature resistivity of 10\textsuperscript{4} $\Omega$m, five to six orders of magnitude lower than single crystal YIG \cite{Thiery2018ElectricalTemperatures,Metselaar1978ElectricalGarnet}.

\begin{figure}[htbp]
    \centering
    \includegraphics[width=0.9\linewidth]{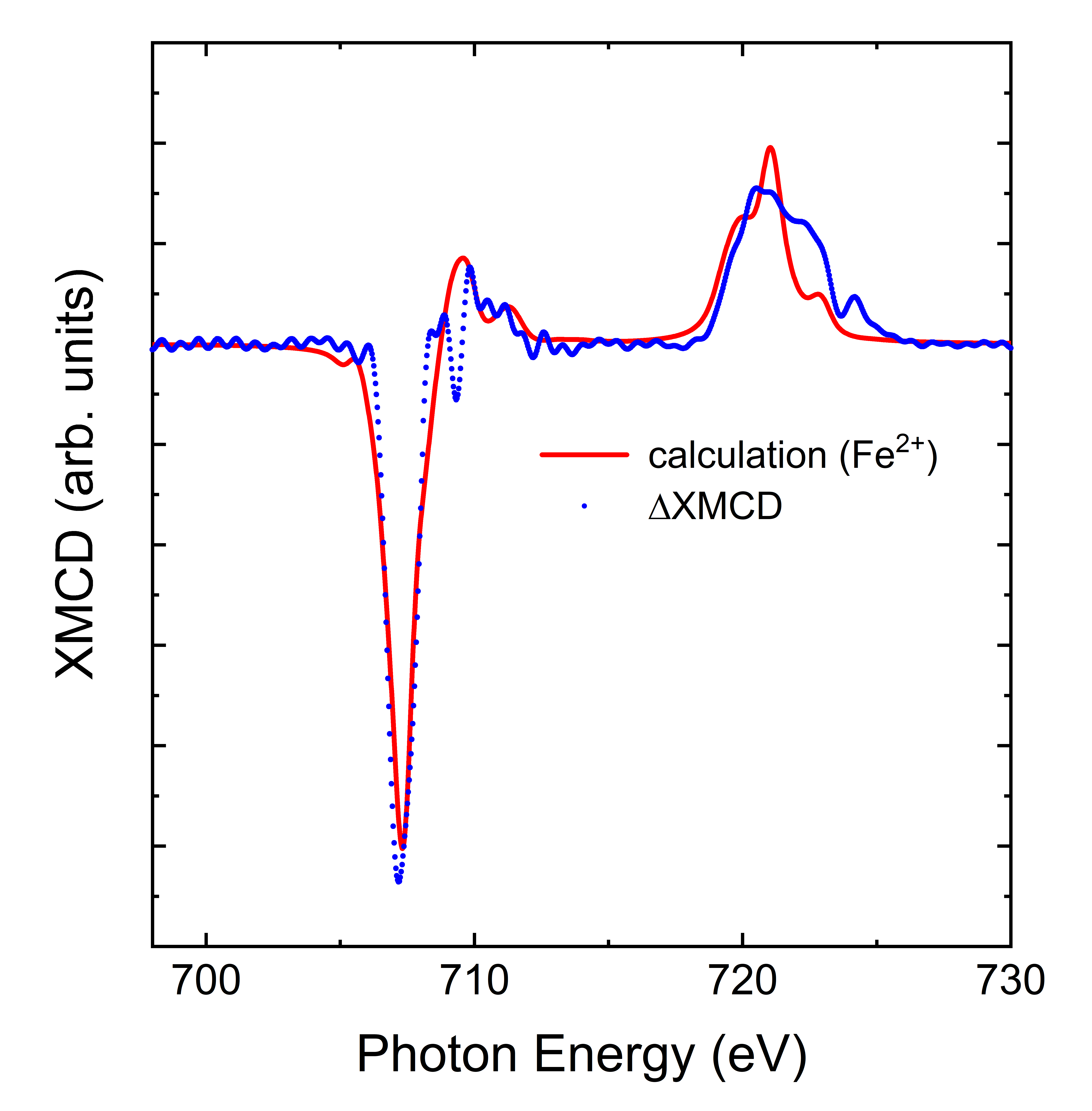}
    \caption{Comparison of $\Delta$XMCD = XMCD(10~K) $-$ XMCD(300~K) for a-YIG/YAG with an atomic multiplet calculation for Fe$^{2+}$ in C$_4$ symmetry ($10Dq = 0.6$ eV, $Ds = 0.15$ eV, $Dt = 0.06$ eV; $d_{z^2}$ ground state orbital). The calculation uses $T = 10$ K, an exchange field of 10 meV, Lorentzian broadening of 0.3 eV (below 715 eV) and 0.4 eV (above), and Gaussian broadening of 0.2 eV.}
    \label{fig:calc}
\end{figure}

Fe$^{2+}$ is plausible in these films: the a-YIG was deposited at room temperature in an oxygen partial pressure (2.5$\times 10^{-3}$ mbar), conditions that may favour oxygen vacancies charge-compensated by Fe$^{2+}$. The calculated a-YIG XMCD does not  match experimental XMCD to the same extent as for the crystalline YIG, even when including a typical $O_h$ Fe$^{2+}$ spectral contribution, see figure~\ref{fig:xmcd300}(b). It is important to emphasize that atomic multiplet calculations are sensitive to the choice of Slater parameters, crystal field values, spin-orbit coupling and charge transfer effects entered into the calculation \cite{vanderLaan2014X-rayMagnetism, XMCD}, and these are likely different between the crystalline and amorphous environments. Secondly, the crystal field calculations are simplified approximations to the actual (semi-)covalent bonds and one would not expect perfect agreement without going into full ab-initio schemes \cite{Figgis}. In particular, these calculations assume an undistorted O$_h$ Fe$^{2+}$ site, whereas Fe$^{2+}$ in an amorphous network is likely to occupy strongly distorted environments; calculations for distorted Fe$^{2+}$ are presented below (figure~\ref{fig:calc}). To constrain the Fe$^{2+}$ content of the films we use XANES measured at the Fe K-edge. We note that TEY probes only the top $\approx$ 6 nm of the sample, whereas K-edge XANES averages over the full 80 nm thickness, so the two techniques need not see the same Fe$^{2+}$ fraction.

\begin{figure}[htbp]
    \centering
    \includegraphics[width=\linewidth]{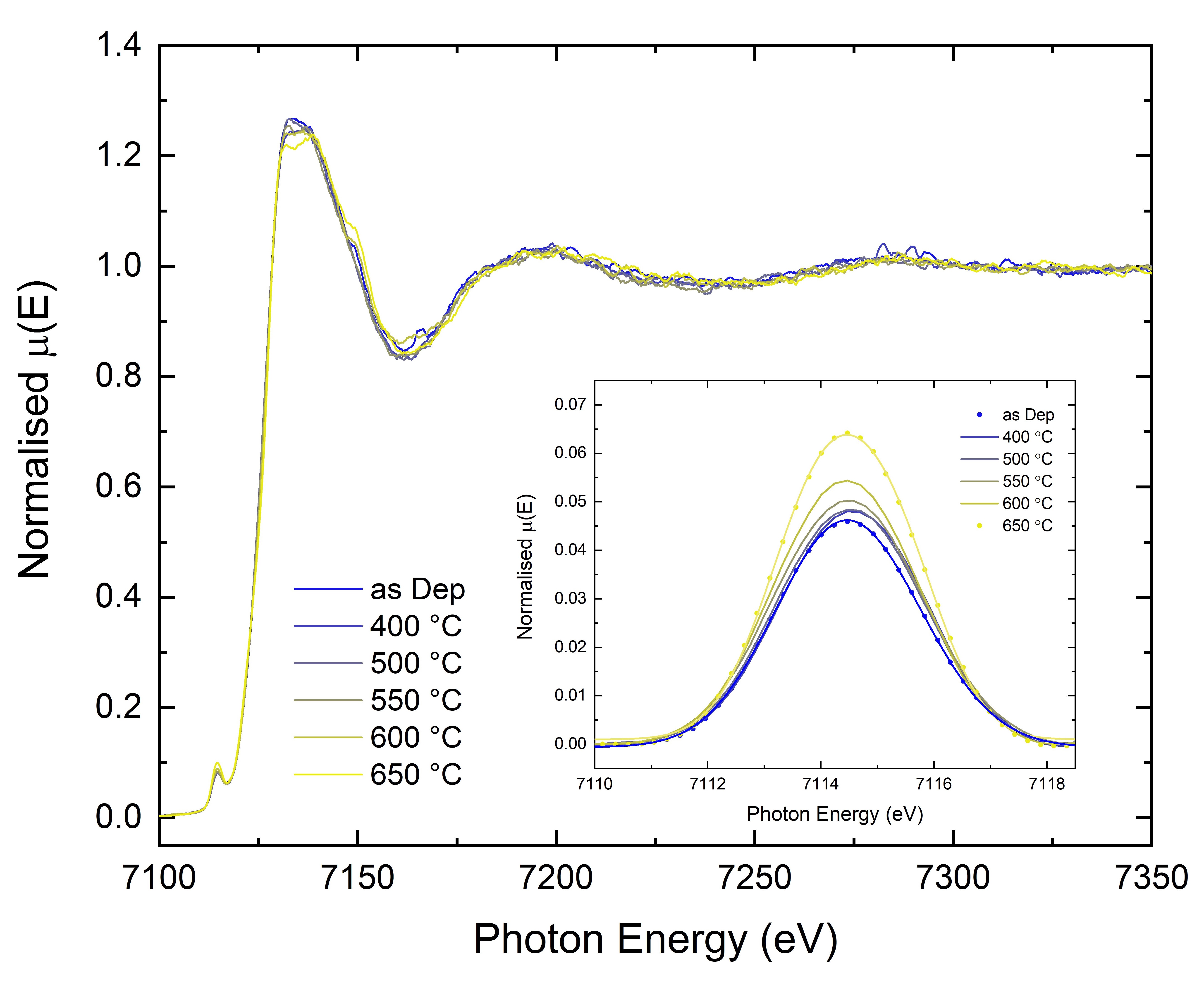}
    \caption{XAS taken over the Fe K-edge for a-YIG/YAG and YIG/YAG films, subject to different ex-situ annealing temperatures. Inset shows pre-edge XANES of the 1s $\rightarrow$ 3d transition, following spline subtraction.}
    \label{fig:xanes}
\end{figure}
XANES data over the Fe K-edge for the a-YIG/YAG thin films, shown in figure~\ref{fig:xanes}, were analysed to determine the dominant Fe valence state by examining the 1s to 3d pre-edge XANES feature, following methods from \cite{Berry2003XANESGlass, Galdenzi2020AccurateK-edge}. The XAS data were normalised at the K-edge step using the ATHENA program \cite{Ravel2005ATHENAIFEFFIT}, and a spline fit was applied to the main absorption edge to isolate the 1s to 3d pre-edge feature. The spectral shape was fit using pseudo-Voigt (p-V) functions to evaluate the relative contributions as described in \cite{GALOISY2001307}. All fits indicated a dominant Fe\textsuperscript{3+} species, with an energy centroid of (7114.4 $\pm$ 0.2) eV. The inset of figure~\ref{fig:xanes} shows the isolated pre-edge feature for all films.  For Fe$^{3+}$ in an octahedral site, two electronic transitions are expected from the 1s core level. These transitions are resolved, provided the
instrumental resolution (including the $\approx$1.1 - 1.2 eV Fe 1s core-hole lifetime broadening) is better than the crystal field splitting, 10Dq. We therefore fit the data with a linear combination of components for O$_{h}$ Fe$^{3+}$ (2 p-V separated by 1.6 eV) and a single p-V for the T$_{d}$ Fe$^{3+}$.    This fitting was consistent for all tested annealing temperatures and comparable XANES data observed in YIG/GGG thin films \cite{Cheshire2022}. In addition, we observe that the weighting of T$_{d}$ to O$_{h}$ components increases by around 20\% when comparing as deposited a-YIG to crystalline YIG, consistent with the XMCD data and the fact that tetrahedral sites give more intense pre-edge features than octahedral; the pre-edge peak increases in magnitude as the annealing temperature is increased.

The lack of spectral features at energies below 7113 eV in the pre-edge peak also indicates a very high Fe\textsuperscript{3+}/$\Sigma$Fe ratio. XANES spectra found in the literature with Fe\textsuperscript{3+}/$\Sigma$Fe = 0.87 show significant skew from 7114 eV towards lower x-ray energies \cite{Berry2003XANESGlass}. Our data therefore limit Fe$^{2+}$ to a minority $\ll$ 13\% of the total Fe when averaged over the film thickness. A minority population at less than a few-per-cent level is not excluded, however, and as shown below it can dominate the temperature-dependent XMCD.
\begin{figure}[htbp]
    \centering
    \includegraphics[width=\linewidth]{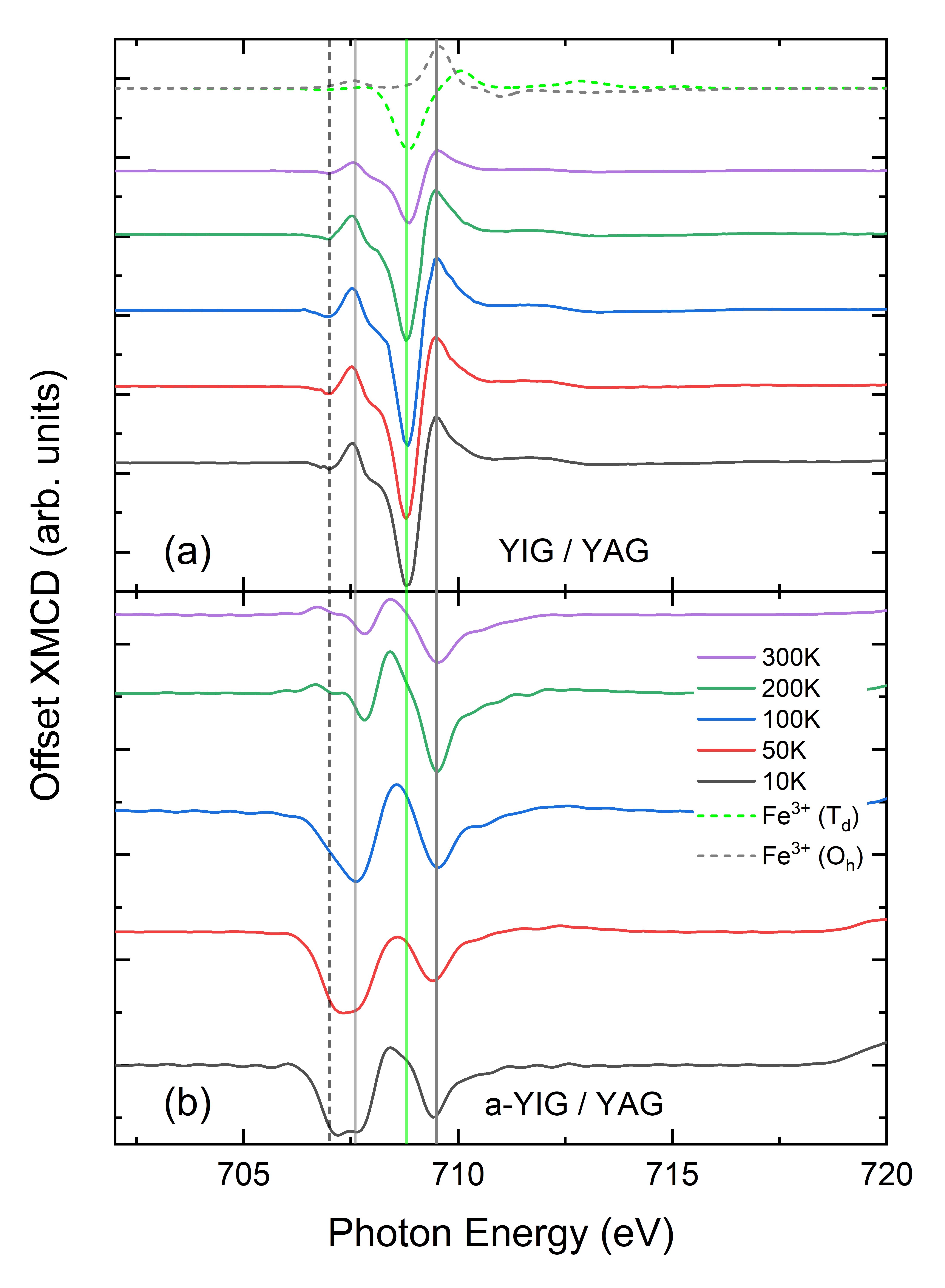}
    \caption{Experimental XMCD spectra of (a) YIG/YAG and (b) a-YIG/YAG for different sample temperatures. Normal Incidence, TEY, B = 5~T. The feature that grows on the low-energy side of the $L_3$ edge on cooling in (b) is discussed in the text.}
    \label{fig:xmcdT}
\end{figure}

\begin{figure}[htbp]
    \centering
    \includegraphics[width=\linewidth]{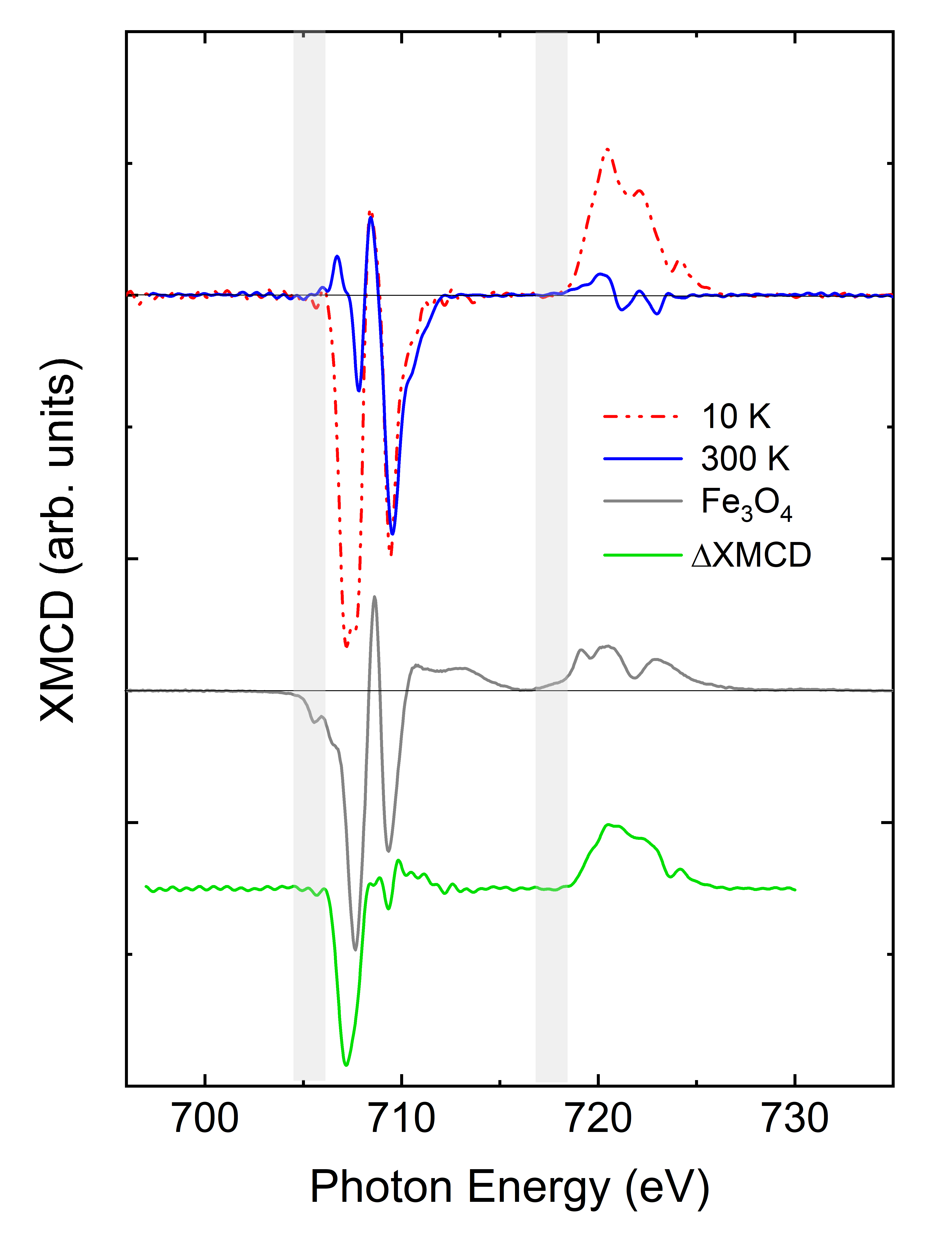}
    \caption{Comparison of experimental XMCD spectra for a-YIG/YAG at 300~K (blue line) and 10~K (red dash-dotted line) with a reference spectrum of magnetite \cite{Love2023}($\text{Fe}_3\text{O}_4$, gray line containing $O_h$ $\text{Fe}^{2+}$). Normal incidence, TEY, $B = 5\text{ T}$. Vertical gray shaded bands mark the characteristic $\text{Fe}^{2+}$ multiplet energy windows at the $L_3$ pre-edge ($705.0$--$706.2\text{ eV}$) and $L_2$ pre-edge ($717.0$--$718.5\text{ eV}$). The green line shows the difference $\Delta$XMCD = XMCD(10~K) $-$ XMCD(300~K), scaled by $\times$0.5.}
    \label{fig:fe3o4}
\end{figure}

Figure~\ref{fig:xmcdT} shows the temperature dependence of the Fe $L_{3}$ XMCD for crystalline YIG/YAG and amorphous a-YIG/YAG. Before the XMCD was calculated, each average ($\sigma^+ + \sigma^-$)/2 spectrum was normalised to the range 0 - 1. This compensates for experimental artefacts, such as ice build-up on the surface during cooling that lowers the XAS intensity, and with it the XMCD magnitude. For crystalline YIG (figure~\ref{fig:xmcdT}(a)), the XMCD lineshape stays the same from 300 to 10~K and scales uniformly with the magnetisation, as expected for a well-ordered ferrimagnet.
 The dominant species that aligns parallel to the applied magnetic field is the $T_d$ $\text{Fe}^{3+}$ sub-lattice (represented by the negative peak at 708.8 eV). In a-YIG however (figure~\ref{fig:xmcdT}(b)), the magnetic alignment is inverted, with the $O_h$ $\text{Fe}^{3+}$ sublattice aligning parallel to the applied field. Intriguingly, the peaks due to $T_d$ $\text{Fe}^{3+}$ and $O_h$ $\text{Fe}^{3+}$ sub-lattices, at similar energies to those found in the crystalline material, are relatively insensitive to temperature. The opposite sign of the $T_d$ peak (aligned antiparallel to the applied field) relative to the $O_h$ peak (parallel to the field) indicates that short-range antiferromagnetic $O_h\text{–}T_d$ superexchange remains active in the amorphous phase, with $O_h$-coordinated $\text{Fe}^{3+}$ dominating the net moment, rather than the sites acting purely as independent paramagnets. This exchange-coupled, weakly temperature-dependent network is distinct from the strongly temperature-dependent feature that emerges at lower energy upon cooling.  As the temperature is reduced to 10~K, the low-energy region of the $L_3$ edge ($706$ - $708\text{ eV}$) undergoes a massive, non-linear intensity enhancement and broadens relative to the initially larger, higher energy O$_h$ peak. At 10 K it becomes apparent that broadening of the peak is due to another spectral feature which starts to become resolved at low temperatures. To identify the origin of the low-temperature dichroic enhancement and test whether it arises from $\text{Fe}^{2+}$, figure~\ref{fig:fe3o4} compares the experimental XMCD of a-YIG at 300~K and 10~K directly against a reference spectrum of magnetite ($\text{Fe}_3\text{O}_4$)\cite{Love2023}, which contains a well-defined $O_h$ $\text{Fe}^{2+}$ component.
Shaded vertical bands highlight the energy windows of the characteristic $\text{Fe}^{2+}$ dichroic signatures of Fe$_3$O$_4$ at the $L_3$ and $L_2$ edges. In $\text{Fe}_3\text{O}_4$, O$_h$ Fe$^{2+}$ produces prominent dichroism across both edges, including low-energy $L_3$ shoulders (705 - 706.5 eV) and an early onset of the positive $L_2$ dichroism, which rises through the pre-$L_2$ window ($717.0$--$718.5\text{ eV}$). The low-temperature a-YIG feature differs from this reference in three respects: it peaks at $\approx$  707.2 eV, $\approx$ 0.4 eV below the main Fe$^{2+}$ XMCD peak of Fe$_3$O$_4$ ($\approx$ 707.6 eV); it lacks the low-energy $L_3$ shoulders; and its $L_2$ onset lies $\approx$1 eV higher ($\approx$ 718.5 eV). These differences are expected if the Fe$^{2+}$ occupies distorted sites with a weaker crystal field than in Fe$_3$O$_4$, as confirmed by the multiplet calculations below. Reducing $10Dq$ compresses the Fe$^{2+}$ multiplet, moving the low-energy structure up towards the main peak and the main peak to slightly lower energy, and similarly raising the onset of the $L_2$ dichroism. The feature also lies below the low-energy edge of the O$_h$ Fe$^{3+}$ multiplet ($\approx$ 707.6--707.8 eV), whereas a reduced crystal field acting on Fe$^{3+}$ would move its low-energy structure upwards rather than downwards. Isolated Fe$^{2+}$ ions would be strongly polarised under these conditions (for $S = 2$, $g \approx 2$ spins in 5~T, the Brillouin function gives $\langle M\rangle/M_{\rm sat} \approx 0.02$ at 300~K and $\approx 0.57$ at 10~K), consistent with the strong, non-linear growth of the feature on cooling. To isolate the additional low-temperature dichroism we plot the difference $\Delta$XMCD = XMCD(10~K) $-$ XMCD(300~K), which shows a negative $L_3$ feature at $\approx$ 707.2 eV and relatively broad $L_{2,3}$ dichroism, consistent with Fe$^{2+}$ in a distribution of distorted environments. Notably, the $L_3$ dichroism in $\Delta$XMCD is concentrated almost entirely in this single low-energy feature, with negligible weight at the T$_d$ (708.8 eV) or O$_h$ (709.5 eV) Fe$^{3+}$ peak positions. A paramagnetic Fe$^{3+}$ population of any coordination would contribute its full multiplet, including strong dichroism near 709--710 eV. Moreover, the main Fe$^{3+}$ XMCD peak moves by only $\approx$ 0.7 eV between O$_h$ and T$_d$ coordination (709.5 to 708.8 eV, figure~\ref{fig:xmcd300}(a)), whereas the $\Delta$XMCD feature lies $\approx$ 1.6 eV below the T$_d$ peak, comparable to the Fe$^{2+}$/Fe$^{3+}$ chemical shift\cite{Miedema2013,vanAken2002}.
To test this interpretation, atomic multiplet calculations were performed for Fe$^{2+}$ using CTM4XAS \cite{Stavitski2010} in C$_4$ symmetry, with the exchange field along the fourfold axis. A cubic crystal field of $10Dq = 0.6$ eV, roughly half the value typical of octahedral Fe$^{2+}$ in oxides, was used to represent the weaker field at under-coordinated sites, and the tetragonal distortion was described by $Ds$ and $Dt$. Slater integrals were reduced to 80\% of their atomic values. A Lorentzian broadening of 0.3 eV below 715 eV and 0.4 eV above, a Gaussian broadening of 0.2 eV, $T = 10$ K and an exchange field $M = 10$ meV were used in the model. Because $M$ fully polarises the ground state, the calculations correspond to the saturated limit and are compared with experiment in lineshape only. All calculated spectra were shifted by $-$1 eV.
In high-spin Fe$^{2+}$ the five majority-spin $3d$ orbitals are filled, so the orbital moment is set by the single minority-spin electron, which occupies the lowest $3d$ orbital. Figure~\ref{fig:calc} compares $\Delta$XMCD with the calculation for $Ds = 0.15$ eV and $Dt = 0.06$ eV, for which the $d_{z^2}$ orbital lies lowest, $\approx$ 0.15 eV below $(d_{xz}, d_{yz})$. The calculation reproduces the principal features of $\Delta$XMCD: a single dominant negative $L_3$ peak, a weak negative feature on its low-energy side, and a strong positive $L_2$ contribution, with an integrated $|L_2/L_3|$ ratio of {0.96} compared with $\approx$ 0.93 for $\Delta$XMCD, characteristic of a quenched orbital moment. After the rigid shift, the calculated $L_3$ peak lies within $\approx$ 0.1 eV of the experimental feature. The calculated $L_2$ structure is sharper than in experiment, as expected for a single site rather than a distribution of distorted environments. Control calculations are shown in the Supplementary Material (figure S2 and table S1). Undistorted O$_h$ Fe$^{2+}$ with the same $10Dq$ gives a split $L_3$ peak and weak $L_2$ dichroism, reflecting an unquenched orbital moment, as do distortions that leave the $(d_{xz}, d_{yz})$ doublet lowest. A compressed site with $d_{xy}$ lowest ($Ds = -0.10$ eV, $Dt = -0.02$ eV) is only partly quenched: the orbital sum rule applied to the calculated spectra gives $\langle \hat{L}_z \rangle = 0.31$, against 0.95 with $(d_{xz}, d_{yz})$ lowest and 0.04 for F2-6, and its $|L_2/L_3|$ ratio of 0.42 remains far from that of $\Delta$XMCD (table S1). A stronger elongation ($Ds = 0.20$ eV, $Dt = 0.10$ eV) gives a quenched spectrum similar to figure~\ref{fig:calc}. Calculations for Fe$^{3+}$ with the same crystal-field parameters (table S2 and figure S3) are spin-only for every distortion, with $|L_2/L_3| \approx 0.99$ and $\langle \hat{L}_z \rangle \leq 0.005$, so the integrated ratio alone does not distinguish the two valences. The lineshape and energy do. With the same rigid energy shift, distorted Fe$^{3+}$ also gives a single dominant negative $L_3$ peak, but $\approx$ 2 eV above the $\Delta$XMCD feature and coincident with the O$_h$ Fe$^{3+}$ peak of crystalline YIG. It is also preceded by a positive feature at $\approx$ 708.0 eV that is absent from $\Delta$XMCD, and its $L_2$ dichroism peaks at $\approx$ 722.5 and 724.3 eV rather than near 721 eV.

\begin{figure}[htbp]
\centering
\includegraphics[width=\linewidth]{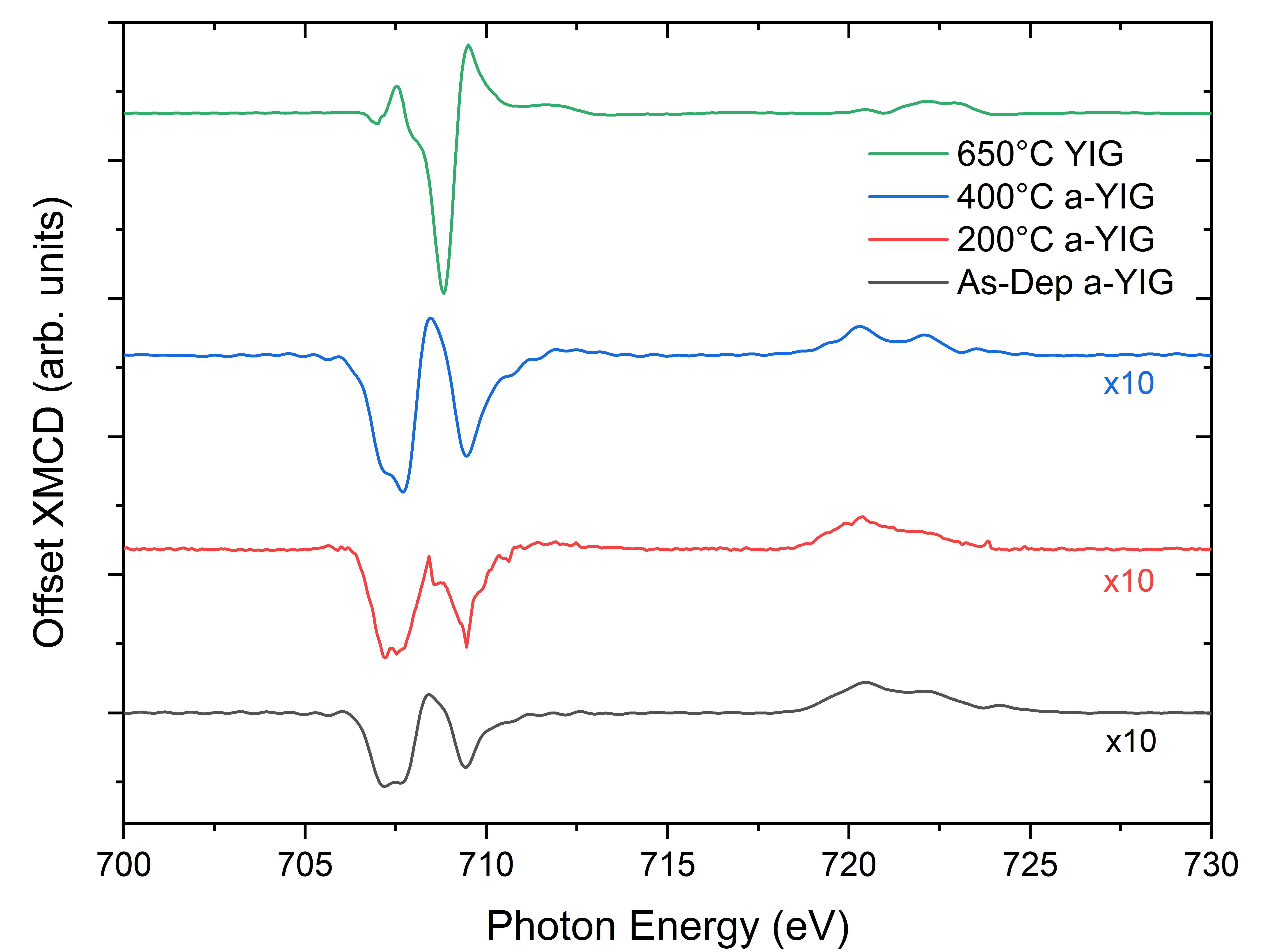}
\caption{Experimental XMCD spectra for a-YIG thin films: as-deposited, or ex-situ annealed at 200~\celsius{} and 400~\celsius{}, compared with crystalline YIG (650~\celsius{}). The a-YIG spectra are multiplied by 10. Measured at T = 10 K.}
\label{fig:anneal}
\end{figure}

Finally, we demonstrate that magnetism in a-YIG remains stable against ex-situ annealing in air across the tested temperature range. Compared to the change upon the a-YIG recrystallising at 650~\celsius{}, magnetism in a-YIG/YAG is essentially unchanged by ex-situ annealing at 200~\celsius{} and 400~\celsius{}, remaining similarly small in all cases (figure~\ref{fig:anneal}).

\section{Discussion}
The low-temperature evolution cannot be explained by a temperature-dependent change in the single-ion multiplet of $\text{Fe}^{3+}$ itself. High-spin $\text{Fe}^{3+}$ ($3d^5$) possesses a spherically symmetric $^6A_1$ ground state with zero orbital angular momentum ($L = 0$). Consequently, structural distortions alter covalent hybridisation and broaden peak profiles, but cannot intrinsically drive significant, temperature-dependent lineshape transformations within a single multiplet. The low-temperature feature must therefore arise from a distinct, magnetically decoupled population of Fe ions whose polarisation depends strongly on temperature.

 XAS difference spectra (see figure S1) between crystalline YIG and a-YIG reveal derivative-like wiggles flanking the main absorption edge rather than a distinct, chemically shifted $\text{Fe}^{2+}$ absorption peak, consistent with inhomogeneous broadening from a distribution of local bond lengths and crystal field parameters ($10Dq$) in the amorphous network. This does not exclude a minority Fe$^{2+}$ population. Every Fe ion contributes to the isotropic XAS, so a few per cent of Fe$^{2+}$ produces only a weak, broadened signal there.
In the XMCD at 10~K and 5~T, by contrast, paramagnetic Fe$^{2+}$ is strongly polarised while the exchange-coupled Fe$^{3+}$ network is largely compensated, so the same minority can dominate the temperature-dependent dichroism.

In undistorted O$_h$ Fe$^{2+}$, as in Fe$_3$O$_4$, the $^5T_{2g}$ ground state retains an unquenched orbital moment, and the $L_2$ dichroism is relatively weak compared with $L_3$ (integrated $|L_2/L_3| \approx 0.31$, table S1). $\Delta$XMCD (figure~\ref{fig:fe3o4}) instead shows a strong positive $L_2$ contribution with $|L_2/L_3|$ close to 1, the value expected for a pure spin moment. For Fe$^{2+}$ this requires the orbital moment to be quenched. A reduced $10Dq$ alone does not achieve this, because the O$_h$ ground state remains an orbital triplet. Which orbital the minority-spin electron occupies under a tetragonal distortion is therefore decisive. Because the majority-spin shell is full, the orbital moment $\langle \hat{L}_z \rangle$ (in units of $\hbar$), where $\hat{L}_z = \sum_i \hat{l}_{z,i}$ is the total orbital angular momentum operator, equals the expectation value of the one-electron operator $\hat{l}_z$ for the minority-spin electron. If the $(d_{xz}, d_{yz})$ doublet lies lowest, as for a moderate elongation, an orbital moment $\langle \hat{L}_z \rangle \approx 1$ survives. If $d_{xy}$ lies lowest, as for a compression, $\hat{l}_z$ couples it only to $d_{x^2-y^2}$, which lies exactly $10Dq$ higher irrespective of $Ds$ and $Dt$. Spin-orbit coupling therefore restores $\langle \hat{L}_z \rangle \approx 4\zeta_{3d}/10Dq \approx 0.35$ for $10Dq = 0.6$ eV ($\zeta_{3d} \approx 0.05$ eV), independent of $Ds$ and $Dt$. The calculations confirm this: the orbital sum rule gives $\langle \hat{L}_z \rangle = 0.31$ and 0.32 for two different compressions (F2-4 and F2-5), falling to 0.16 when $10Dq$ is doubled (F2-9). Only when $d_{z^2}$ lies lowest, which requires a strong axial elongation ($Ds + 10Dt > 10Dq$), does $\hat{l}_z$ have no matrix elements to the other $3d$ orbitals, so that the orbital moment is quenched along the distortion axis. For field directions perpendicular to the axis, spin--orbit coupling to $(d_{xz}, d_{yz})$ leaves a small residual moment, so a randomly oriented ensemble gives a ratio close to, but not necessarily equal to, unity. The $|L_2/L_3|$ ratio therefore does not by itself distinguish quenched Fe$^{2+}$ from Fe$^{3+}$; the assignment rests instead on the energy of the feature, the absence of $\Delta$XMCD weight at the Fe$^{3+}$ peak positions, and its Curie-like temperature dependence.

We therefore attribute the strong temperature dependence to a Curie-paramagnetic sub-population of magnetically isolated Fe$^{2+}$ ions at severely distorted, under-coordinated sites. Room-temperature growth at moderate oxygen partial pressure is expected to leave some oxygen vacancies in the amorphous network, which are charge-compensated by Fe$^{2+}$. The missing oxygen both severs Fe--O--Fe superexchange pathways, decoupling these ions from the Fe$^{3+}$ network, and lowers the local coordination and symmetry. Removal of an apical oxygen leaves a strongly axially elongated five-coordinate site which produces the distortion and weak average crystal field inferred from the calculations, and places $d_{z^2}$ lowest. At 300~K, thermal energy randomises these uncoupled spins; on cooling to 10~K, the applied 5~T field aligns them, producing a localized, uncompensated, strongly temperature-dependent dichroic signal on the low-energy side of the bulk-like O$_h$ Fe$^{3+}$ multiplet. The distortion therefore sets the energy and orbital character of the feature, while its temperature dependence arises from Brillouin polarisation. Because these Fe$^{2+}$ ions are polarised far more efficiently than the exchange-coupled, largely compensated Fe$^{3+}$ network, a population below the XANES detection limit can account for the signal.
The persistence of the low-temperature feature after annealing in air at 200 and 400~\celsius{} (figure~\ref{fig:anneal}) indicates that these Fe$^{2+}$ sites are not readily re-oxidised, suggesting that they are stabilised within the amorphous network, for example by bound oxygen vacancies. Fe$^{2+}$ adjacent to O$_h$ Fe$^{3+}$ would also provide the Fe$^{2+}$--Fe$^{3+}$ hopping pathway discussed above, offering a microscopic link between the defect chemistry of a-YIG, its low resistivity and the charge-transfer contribution to non-local transport.

\section{Conclusion}
In conclusion, we have demonstrated that the Fe environment in amorphous YIG differs fundamentally from its crystalline counterpart. XMCD over the Fe $L_{2,3}$ edges shows that a-YIG magnetism is an order of magnitude weaker than crystalline YIG and is dominated by $O_h$-coordinated $\text{Fe}^{3+}$ species. O\textsubscript{h}-dominant contributions to a-YIG magnetism are seen to be stable with ex-situ annealing up to at least 400~\celsius{} consistent with \cite{Yu2024}. On cooling, an additional, strongly temperature-dependent dichroic feature appears at $\approx$ 707.2 eV. Comparison with Fe$_3$O$_4$ shows that it does not arise from Fe$^{2+}$ in undistorted O$_h$ sites. Instead, its energy, the absence of accompanying dichroism at the Fe$^{3+}$ peak positions, its Curie-like temperature dependence and near spin-only $L_2/L_3$ ratio are consistent with a minority population of magnetically decoupled Fe$^{2+}$ ions at strongly distorted, under-coordinated sites, where a reduced crystal field compresses the multiplet and a strong axial distortion places the minority-spin electron in a $d_{z^2}$ orbital, quenching the orbital moment. Atomic multiplet calculations for Fe$^{2+}$ in this geometry reproduce the lineshape of the low-temperature signal, whereas undistorted or compressed Fe$^{2+}$ sites do not. Bulk Fe $K$-edge XANES, with Fe$^{3+}$/$\Sigma$Fe close to 1, limits this population to a small fraction of the total Fe, which nonetheless dominates the low-temperature XMCD because it is polarised far more efficiently than the largely compensated Fe$^{3+}$ network. Such Fe$^{2+}$ sites, associated with oxygen vacancies, offer a microscopic origin for hopping conduction in a-YIG and may contribute to the sample-to-sample variation in its reported spin-transport properties.

\begin{acknowledgments}
The authors acknowledge Diamond Light Source for time on Beamline I10 under proposal MM22550-1, and Beamline B18 under proposal SP24930-1.
\end{acknowledgments}

\bibliographystyle{apsrev4-2}
\bibliography{YIG3}

\clearpage
\onecolumngrid
\begin{center}
\textbf{\large Supplementary Material}
\end{center}
\setcounter{section}{0}\setcounter{figure}{0}\setcounter{table}{0}\setcounter{equation}{0}
\renewcommand{\thesection}{S\arabic{section}}
\renewcommand{\thefigure}{S\arabic{figure}}
\renewcommand{\thetable}{S\arabic{table}}
\renewcommand{\theequation}{S\arabic{equation}}

\section{XAS difference spectra}

\begin{figure}[htbp]
\centering
\includegraphics[width=0.7\linewidth]{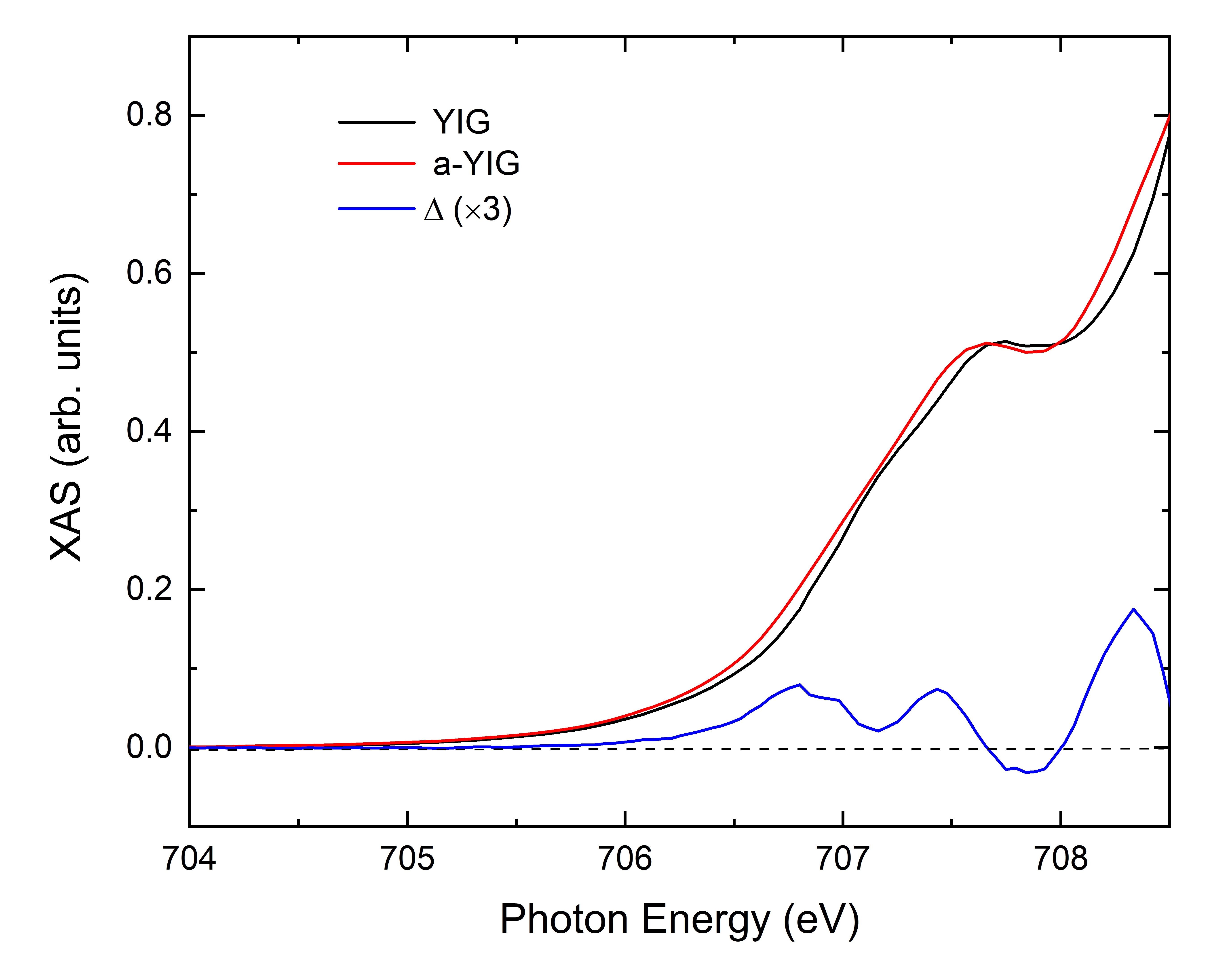}
\caption{Low energy region of average XAS for YIG (black) and a-YIG (red) measured at $T = 10$~K. The spectra have been normalised to the $L_3$ edge shoulder. The difference spectrum, $I_{\mathrm{a\text{-}YIG}} - I_{\mathrm{YIG}}$ (multiplied by 3), is shown in blue.}
\label{fig:S1}
\end{figure}

\section{Atomic multiplet calculations for Fe$^{2+}$}

All calculations used CTM4XAS \cite{Stavitski2010} in C$_4$ symmetry with the exchange field along the fourfold axis.
Common parameters: $10Dq = 0.6$~eV (except F2-9), Slater integrals at 80\% of their atomic values,
 exchange field $M = 10$~meV, $T = 10$~K,
Lorentzian broadening of 0.3~eV below 715~eV and 0.4~eV above, and Gaussian broadening of 0.2~eV. The photon energy axis in CTM4XAS is not absolute; the calculated spectra therefore require an energy shift to match the experimental energy scale, but the energy difference between the main $L_3$ peak in Fe$^{2+}$ and Fe$^{3+}$ is found experimentally to be $\approx 1.5$--1.7~eV \cite{Miedema2013,vanAken2002}.
The XMCD is the right- minus left-circularly polarised spectrum. The integrals
\begin{equation}
p = \int_{L_3} \mathrm{XMCD}\,dE, \qquad q = \int_{L_3 + L_2} \mathrm{XMCD}\,dE
\end{equation}
give $q/p \approx 0$ for a quenched orbital moment. Because $p$ also depends on the spin sum rule, including the magnetic dipole term $\langle T_z \rangle$, $q/p$ is not a like-for-like measure of the orbital moment between different ground states. The orbital moment is therefore also given from the orbital sum rule alone,
\begin{equation}
\langle \hat{L}_z \rangle = \frac{4}{3}\, n_h \frac{|q|}{r}, \qquad r = \int_{L_3 + L_2} (I_{\mathrm{left}} + I_{\mathrm{right}})\,dE,
\end{equation}
with $n_h = 4, 5$ for Fe$^{2+,\,3+}$ respectively, and $|L_2/L_3| = {|q-p|}\, /\,{|p|}$.

\begin{table*}[htbp]
\centering
\caption{Crystal-field parameters, lowest one-electron orbital of the minority-spin electron, integrated XMCD and orbital moment (orbital sum rule) for the Fe$^{2+}$ calculations.}
\label{tab:fe2}
\begin{tabular}{lccclccc}
\toprule
Run & $10Dq$ (eV) & $Ds$ (eV) & $Dt$ (eV) & Lowest orbital & $q/p$ & $|L_2/L_3|$ & $\langle \hat{L}_z \rangle$ \\
\midrule
F2-1 & 0.6 & 0     & 0     & $t_{2g}$ (cubic)   & 0.687 & 0.31 & 0.76 \\
F2-2 & 0.6 & 0.10  & 0.02  & $d_{xz}, d_{yz}$   & 0.626 & 0.37 & 0.95 \\
F2-3 & 0.6 & $-$0.02 & $-$0.10 & $d_{xz}, d_{yz}$ & 0.613 & 0.39 & 0.95 \\
F2-4 & 0.6 & $-$0.10 & $-$0.02 & $d_{xy}$       & 0.584 & 0.42 & 0.31 \\
F2-5 & 0.6 & $-$0.20 & $-$0.05 & $d_{xy}$       & 0.607 & 0.39 & 0.32 \\
F2-6 & 0.6 & 0.15  & 0.06  & $d_{z^2}$          & 0.045 & 0.96 & 0.04 \\
F2-7 & 0.6 & 0.20  & 0.10  & $d_{z^2}$          & 0.006 & 0.99 & 0.01 \\
F2-8 & 0.6 & 0.30  & 0.10  & $d_{z^2}$          & 0.006 & 0.99 & 0.01 \\
F2-9 & 1.2 & $-$0.10 & $-$0.02 & $d_{xy}$       & 0.372 & 0.63 & 0.16 \\
\bottomrule
\end{tabular}
\end{table*}

\begin{figure}[htbp]
\centering
\includegraphics[width=0.7\linewidth]{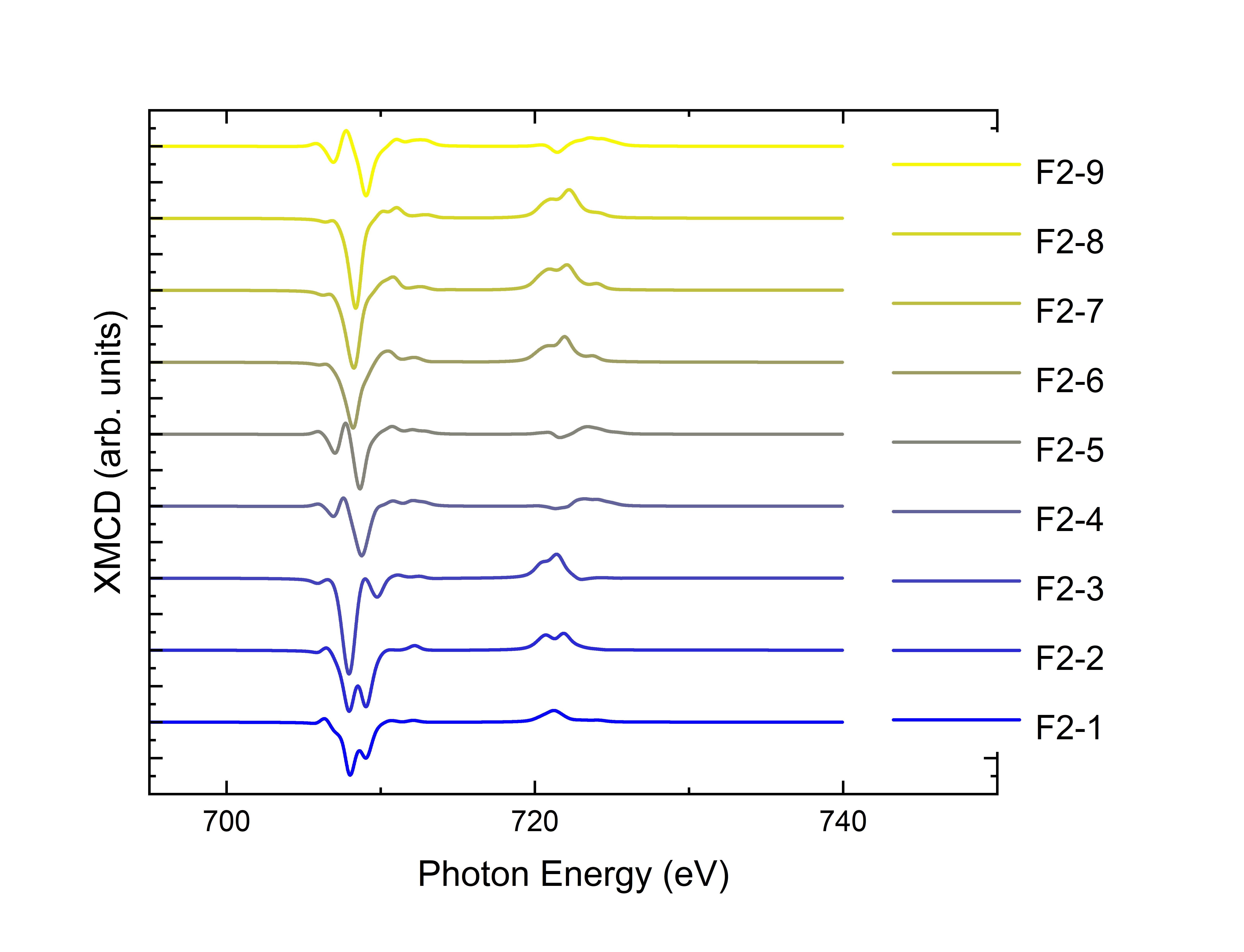}
\caption{Calculated Fe $L_{2,3}$ XMCD of Fe$^{2+}$ for the parameters in table~\ref{tab:fe2}, offset vertically for clarity.
F2-6 is the calculation compared with experiment in the main article.}
\label{fig:S2}
\end{figure}

\clearpage

\section{Atomic multiplet calculations for Fe$^{3+}$}

The Fe$^{3+}$ calculations used the same settings and the same distortions as for Fe$^{2+}$,
together with the octahedral ($10Dq = 1.6$~eV) and tetrahedral ($10Dq = -0.6$~eV) parameters
used for crystalline YIG in the main article.

\begin{table*}[htbp]
\centering
\caption{Crystal-field parameters, integrated XMCD and orbital moment (orbital sum rule) for the Fe$^{3+}$ calculations.}
\label{tab:fe3}
\begin{tabular}{lccclccc}
\toprule
Run & $10Dq$ (eV) & $Ds$ (eV) & $Dt$ (eV) & Comment & $q/p$ & $|L_2/L_3|$ & $\langle \hat{L}_z \rangle$ \\
\midrule
F3-1  & 0.6    & 0       & 0       & cubic, weak field     & 0.0061 & 0.994 & 0.0046 \\
F3-2  & 0.6    & 0.10    & 0.02    & as F2-2               & 0.0061 & 0.994 & 0.0047 \\
F3-3  & 0.6    & $-$0.02 & $-$0.10 & as F2-3               & 0.0061 & 0.994 & 0.0046 \\
F3-4  & 0.6    & $-$0.10 & $-$0.02 & as F2-4               & 0.0058 & 0.994 & 0.0044 \\
F3-5  & 0.6    & $-$0.20 & $-$0.05 & as F2-5               & 0.0056 & 0.994 & 0.0042 \\
F3-6  & 0.6    & 0.15    & 0.06    & as F2-6               & 0.0063 & 0.994 & 0.0048 \\
F3-7  & 0.6    & 0.20    & 0.10    & as F2-7               & 0.0065 & 0.993 & 0.0050 \\
F3-8  & 0.6    & 0.30    & 0.10    & as F2-8               & 0.0065 & 0.993 & 0.0051 \\
F3-9  & 1.6    & 0       & 0       & octahedral YIG site   & 0.0006 & 0.999 & 0.0004 \\
F3-10 & $-$0.6 & 0       & 0       & tetrahedral YIG site  & 0.0062 & 0.994 & 0.0046 \\
\bottomrule
\end{tabular}
\end{table*}

\begin{figure}[htbp]
\centering
\includegraphics[width=0.7\linewidth]{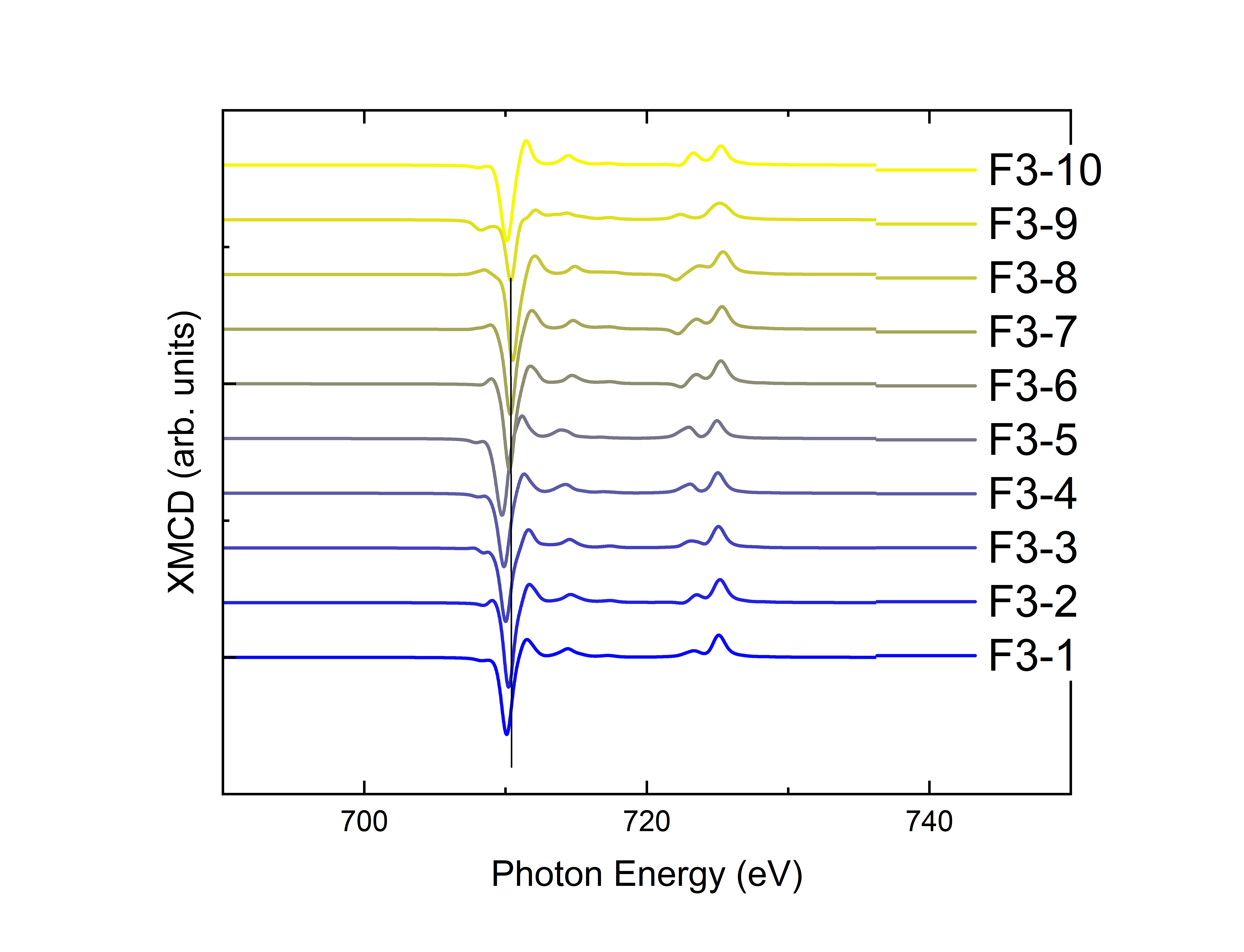}
\caption{Calculated Fe $L_{2,3}$ XMCD of Fe$^{3+}$ for the parameters in table~\ref{tab:fe3}, offset vertically for clarity.}
\label{fig:fe3}
\end{figure}

\end{document}